# YouTube Videos for Public Health Literacy? A Machine Learning Pipeline to Curate Covid-19 Videos

Yawen GUO[a], Xiao LIU[b], Anjana SUSARLA[c], Rema PADMAN[d]
[a] *University of California, Irvine*
[b] *Arizona State University*
[c] *Michigan State University*
[d] *Carnegie Mellon University*

**Abstract.** The COVID-19 pandemic has highlighted the dire necessity to improve public health literacy for societal resilience. YouTube, the largest video-sharing social media platform, provides a vast repository of user-generated health information in a multi-media-rich format which may be easier for the public to understand and use if major concerns about content quality and accuracy are addressed. This study develops an automated solution to identify, retrieve and shortlist medically relevant and understandable YouTube videos that domain experts can subsequently review and recommend for disseminating and educating the public on the COVID-19 pandemic and similar public health outbreaks. Our approach leverages domain knowledge from human experts and machine learning and natural language processing methods to provide a scalable, replicable, and generalizable approach that can also be applied to enhance the management of many health conditions.

## 1. Introduction

The easy availability of vast amounts of medical information on the Internet, coupled with the rapid growth in the use of social media by the public, patients, and clinicians alike, has transformed how consumers access medical information to manage their health information needs and illnesses [1]. Traditionally, patients receive such information and instructions in text format from their healthcare providers and organizations. However, purely text-based medical information has been found to reduce user attention, understanding, recall, and compliance, especially for patients with low literacy levels [2]. Hence, it is important to design educational materials for patients and the general public that increase engagement and participation in health-related decision-making.

YouTube hosts millions of health-related videos about the pathogenesis, diagnosis, treatment, and prevention of many medical conditions, including COVID-19. This vast repository of audio-visual content, created by both professional healthcare organizations and random individuals, is of widely varying quality and a challenge for both the public and healthcare professionals to search and retrieve credible and relevant videos as a just-in-time, contextualized, prescriptive, digital therapeutic and educational intervention.

This study aims to address the challenge of automating the identification, retrieval and curation of useful health-related YouTube video content via a machine learning pipeline from the perspective of improving public health literacy, using COVID-19 as an illustrative example. Evaluating encoded medical content in a video and its understandability are two critical criteria we apply to assess a recommended video [1]. Studies assessing the readability, relevance, or comprehensibility of patient education materials on a myriad of topics abound. The evidence is clear and consistent that most education materials are too complex for most consumers because many adults lack the requisite skills to obtain and process basic health information and services needed to make appropriate health decisions [3]. Responding to the health communication challenges raised by the pandemic, public health agencies must use resources such as YouTube to better deliver timely and accurate information and to minimize the spread of misinformation.

## 2. Methods

### *2.1. Data Collection & Annotation*

To assess the amount of medical information encoded in YouTube videos and their understandability, we develop a data collection process that first generates keywords about COVID-19 and then retrieves the videos. We select the search keywords from three sources including terms extracted from the posts and recommendations of the Expert Answers forum on DailyStrength, search suggestions appearing below the YouTube search bar with the search keyword 'COVID-19', and frequently asked questions (FAQ) lists of the World Health Organization (WHO) and the Centers for Disease Control and Prevention (CDC). Result from the union of these sources, their popularity ranking and selection of top 40 keywords covering users' most searched topics are shown in Table 1.

**Table 1.** Collected Keywords

| |
|---|
| COVID-19, COVID-19 prep, COVID-19 transmission, COVID-19 prevention, sars-cov-2, COVID-19 diagnosis, COVID-19 health, COVID-19 treatment, COVID-19 public health, COVID-19 affect, COVID-19 mental health, COVID-19 recovery, COVID-19 community health, COVID-19 tips, COVID-19 educational videos, COVID-19 impact, COVID-19 contact tracing, COVID-19 supplement, COVID-19 community education, COVID-19 anxiety, COVID-19 epidemiology, COVID-19 explained, COVID-19 wellbeing, COVID-19 animation, COVID-19 infodemiology, COVID-19 children, COVID-19 clinical, COVID-19 outbreak, COVID-19 higher risk population, COVID-19 vaccine, COVID-19 symptoms, COVID-19 test, COVID-19 emergency signs, COVID-19 .edu, COVID-19 experience, COVID-19 rapid test, COVID-19 update, COVID-19 serious patients, COVID-19 education, COVID-19 severe illness |

We collected the top 25 videos per keyword resulting from each search on YouTube in November 2020. Unique video IDs were retained after removing duplicates (different keywords may generate the same videos). We further filter videos by restricting the video language to be English and video duration to be between 60 seconds to 6 minutes for easy viewing by the public. Some videos could not be downloaded because of video license issues. The videos in the final collection were annotated by one author (YG) and three graduate research associates, two of whom were trained in clinical medicine and each annotated half the video set from the domain expert perspective; YG and the other research associate in information technology and analytics annotated them from a consumer perspective. Each had access only to the actual videos and watched the video content and no other information on the YouTube webpage. They annotated the videos according to the following criteria: medical information, understandability, and overall recommendation. For medical information, videos were annotated as encoding high or low levels of medical information on COVID-19 and infectious diseases. Understandability was rated according to the PEMAT framework for audio-visual materials, described later [4]. We also label the videos based on whether the annotator would recommend the video to patients and the public from a consumer perspective and a domain expert perspective, respectively. Finally, we retained the dataset that included videos where both consumers agreed on the labels for overall recommendation score and understandability score, and used experts' annotation scores to resolve inconsistencies when the consumers disagreed. We retained the experts' annotation on medical information since their judgments on this criterion are more accurate and credible.

### *2.2. Video Feature Extraction*

We utilized the YouTube Data API to retrieve video metadata, including channel ID (account name), publishing time of the video, title, description, tags, duration, and definition. Several functions of the Google Cloud Video Intelligence API were applied to detect important features of each video from the perspective of understandability. The Speech-to-Text API transcribes the spoken word in audio format in a video/video segment into text and returns blocks of text for each portion of transcribed audio for analysis. The volume of medical information encoded in the videos is obtained by utilizing the Unified Medical Language System (UMLS) as a reference for medical terminology annotation, with relevant semantic types selected according to existing literature. Results are used to prepare the training texts for the Bidirectional Long Short-Term Memory Recurrent Neural Network (BLSTM-RNN) medical entity recognizer. We utilize a pre-trained model from our prior work to recognize and extract medical terms in the video text data [1] and enhance this extraction process by adding a dictionary method to detect COVID-19 related information in the video data from the Oxford English Dictionary.

PEMAT assesses video understandability based on content completeness, word choice and style, organization of the material, layout and design and the use of visual aids. The understandability score in PEMAT is calculated by adding the scores for each topic divided by total points possible (excluding the N/As) and multiplying by 100 to get the percentage. The higher the score, the more understandable is the material. Annotators rated the collected videos for understandability by following the steps of PEMAT Tool for Audiovisual Materials (PEMAT-A/V) for calculating video understandability score and using a threshold of 50%.

The ranking system of YouTube emphasizes features representing video popularity, such as view count and total watch time. However, when users search on YouTube, the videos with high popularity may not necessarily have high relevance to their search query. Therefore, to evaluate the relevance of the videos collected from the YouTube platform, we calculate 3 cosine similarity metrics between the search keyword and the corresponding title, description, and transcription of each video, respectively.

### *2.3. Classification*

We performed correlation-based feature selection method to reduce feature redundancy and trained several commonly used machine learning algorithms using these features to classify videos into those recommended or not. We focus on the relationship between encoded medical information in videos and their understandability and the video recommendation label from both expert and consumer perspectives.

## 3. Results

The top 25 returned videos for each of the 40 collected search terms were stored alongside their metadata and rankings in the retrieved list. 828 unique video IDs were retained after removing duplicates, of which 549 were downloaded with permission. Applying restrictions on language and video duration resulted in 304 videos being eventually gathered for annotation and classification. Removing features with correlation less than 0.5 led to retention of 20 important features for training the model (Table 2).

**Table 2.** Descriptive summary of selected features

| Feature | Mean | Stdev | Feature | Mean | Stdev |
|---|---|---|---|---|---|
| medically information expert score | 0.47 | 0.5 | # of summary words in description | 0.01 | 0.09 |
| OCR confidence score | 0.91 | 0.09 | like rate | 0.14 | 0.15 |
| # of active verbs in transcription | 0.38 | 0.22 | dislike rate | 0.08 | 0.13 |
| readability score of transcription | 0.19 | 0.09 | # of comments | 0.03 | 0.09 |
| # of shots in the video | 0.22 | 0.19 | # of channel subscribers | 0.06 | 0.11 |
| transcription confidence score | 0.85 | 0.16 | # of channel views | 0.1 | 0.17 |
| # of summary words in transcription | 0.08 | 0.26 | cosine similarity between video description and search keyword | 0.24 | 0.15 |
| cosine similarity between video transcription and search keyword | 0.44 | 0.18 | cosine similarity between video title and search keyword | 0.43 | 0.29 |
| readability score of description | 0.4 | 0.17 | if the video has a description | 0.99 | 0.14 |
| # of sentences in description | 0.11 | 0.12 | understandability | 0.91 | 0.3 |

Logistic regression, support vector machine, random forest and XGBoost were trained to classify the videos on an almost balanced dataset (40% videos recommended by experts) of 243 randomly selected videos and evaluated on the remaining 61 videos using an 80-20 split. Table 3 reports on the model performance, with logistic regression showing the best performance with an overall accuracy of 83.61%. The five most important features identified by the logistic regression classifier included understandability, medical information expert score, readability score of description, number of shots in the video, and cosine similarity between video description and keyword. Table 4 lists examples and descriptions of videos that include both recommended and not recommended videos.

**Table 3.** Classification Performance.

| Classifier | Logistic Regression | SVM | Random Forest | XGBoost |
|---|---|---|---|---|
| **Accuracy** | 83.61% | 77.63% | 83.61% | 73.68% |
| **Recall** | 75.86% | 66.67% | 66.67% | 63.33% |

**Table 4.** Sample videos from the best classifier.

| Recommended | Description | Not Recommended | Description |
|---|---|---|---|
| Transmission of COVID-19 | Factual and timely guidance for combating COVID-19 provided by two doctors, presented with slides | Battleground states see sharp increases in COVID-19 cases before Election Day | News, minimum medical information presented in the video |
| COVID-19 Preventing transmission - Environmental cleaning | Educational video from a university with detailed information on Covid-19 prevention | CDC abruptly reverses guidance on COVID-19 airborne transmission | News, low medical information presented in the video |

## 4. Discussion

The purpose of this study was to retrieve, curate and recommend COVID-19 related YouTube videos with high volume of medical information and high understandability. The results of our ML approach offer several implications for researchers, public health practitioners, as well as online video content creators. Content creators should use auxiliary and supportive means to verify that published content is valid and easy to understand. Some videos may contain good medical content but may be incomprehensible for majority of viewers. Others are well-organized and formatted but are advertisements from clinics or are news updates on the latest information about the positive COVID-19 cases. Some videos do not have audio but have well-organized text or slides showing on the screen, while others have no speech, just videos and images, which are encoded with low-level of medical information, but patients may find them very useful. For example, when patients share their experiences regarding rapid testing or emotional and physical changes after testing positive on COVID-19, other patients may find these videos very helpful for understanding and accepting their own condition. This significant variability poses major challenges in automating the curation process.

This study has some limitations that suggest directions for future research such as addressing potential representativeness bias due to the significant volume of data loss from the original videoID collection to the final classifier dataset. Additional video features that may help delineate specific infectious disease related topics, improve classification performance using deep learning models, new criteria such as content timeliness and accuracy, and evaluation by clinical experts and consumers via randomized experiments and observational studies are possible future extensions.

## 5. Conclusion

In this study, we synthesize information retrieval, deep learning and statistical methods to identify understandable and medically relevant YouTube videos on COVID-19 for dissemination to the public and build the scientific evidence-base for improving societal health and resilience. Our findings are generalizable for other health conditions and have implications for healthcare professionals, patients and educators, information systems researchers, and policymakers.

## References

[1] Liu X, Zhang B, Susarlia A, Padman R. Go to you tube and call me in the morning: Use of social media for chronic conditions. MIS Quarterly. 2020;44(1):257–83.

[2] Drozd B, Couvillon E, Suarez A. Medical youtube videos and methods of evaluation: Literature review. JMIR Medical Education. 2018;4(1).

[3] Tom K, Phang PT. Effectiveness of the video medium to supplement preoperative patient education: A systematic review of the literature. Patient Education and Counseling. 2022;105(7):1878–87.

[4] Shoemaker SJ, Wolf MS, Brach C. Development of the patient education materials assessment tool (PEMAT): A new measure of understandability and actionability for print and audiovisual patient information. Patient Education and Counseling. 2014;96(3):395–403.